\documentclass[
]{ceurart}

\usepackage{listings}
\newtheorem{definition}{Definition}[section]

\newcommand{\Sys}{\mathit{S}}
\newcommand{\probP}{\text{I\kern-0.15em P}}

\begin{document}

\copyrightyear{2024}
\copyrightclause{Copyright for this paper by its authors. Use permitted under Creative Commons License Attribution 4.0 International (CC BY 4.0).}

\conference{AEQUITAS 2024: Workshop on Fairness and Bias in AI | co-located with ECAI 2024, Santiago de Compostela, Spain}

\title{Using Protected Attributes to Consider Fairness in Multi-Agent Systems}

% \tnotetext[1]{You can use this document as the template for preparing your
%   publication. We recommend using the latest version of the ceurart style.}

\author[1]{Gabriele {La Malfa}}[%
email=gabriele.la_malfa@kcl.ac.uk,
]
\cormark[1]
\address[1]{UKRI Centre for Doctoral Training in Safe and Trusted AI, King’s College London, London, WC2B 4BG}

\author[1]{Jie M. {Zhang}}[%
email=jie.zhang@kcl.ac.uk,
]
% \cormark[1]
% \address[1]{King’s College London, London, WC2B 4BG}

\author[2]{Michael {Luck}}[%
email=michael.luck@sussex.ac.uk,
]
% \cormark[1]
\address[2]{University of Sussex, Brighton, BN1 9RH}

\author[1]{Elizabeth {Black}}[%
email=elizabeth.black@kcl.ac.uk,
]
% \cormark[1]
% \address[1]{King’s College London, London, WC2B 4BG}

\cortext[1]{Corresponding author.}
% \fntext[1]{These authors contributed equally.}

\begin{abstract}
Fairness in Multi-Agent Systems (MAS) has been extensively studied, particularly in reward distribution among agents in scenarios such as goods allocation, resource division, lotteries, and bargaining systems. Fairness in MAS depends on various factors, including the system's governing rules, the behaviour of the agents, and their characteristics.
Yet, fairness in human society often involves evaluating disparities between disadvantaged and privileged groups, guided by principles of Equality, Diversity, and Inclusion (EDI). Taking inspiration from the work on algorithmic fairness, which addresses bias in machine learning-based decision-making, we define \emph{protected attributes} for MAS as characteristics that should not disadvantage an agent in terms of its expected rewards.
We adapt fairness metrics from the algorithmic fairness literature—namely, \emph{demographic parity}, \emph{counterfactual fairness}, and \emph{conditional statistical parity}—to the multi-agent setting, where self-interested agents interact within an environment.
These metrics allow us to evaluate the fairness of MAS, with the ultimate aim of designing MAS that do not disadvantage agents based on protected attributes.
\end{abstract}

\begin{keywords}
  Fairness \sep
 bias \sep
  Multi-Agent Systems (MAS)
\end{keywords}

\maketitle

\section{Introduction}
Multi-Agent Systems (MAS) consist of agents interacting with each other and their surrounding environment to achieve their individual or shared goals. 
The achievement of an agent's goals may depend on the actions it takes, the actions of other agents, the environment they are situated in, and the rules that govern the MAS. Similarly, fairness in MAS depends on multiple factors. Fairness can be influenced by agents' decision-making processes, as evidenced by research in reinforcement learning focused on developing fair and efficient policies~\cite{Gajane2022}. It can also hinge on mechanism design, as seen in scenarios like goods allocation games~\cite{Amantidis2023} or cake-cutting problems~\cite{Procaccia2013}, where rules can ensure fair reward distribution among agents. Additionally, fairness can be affected by things like an agent's utility~\cite{Ulle2004, Bertsimas2011} or their priority in accessing resources~\cite{DeJong2008B, Bu2023}, among others.

In human societies, fairness is often defined in terms of characteristics that should not disadvantage an individual or group, such as age, race, disability or gender. For example, in the UK Equality Act 2010\footnote{ ~\hyperlink{https://www.gov.uk/guidance/equality-act-2010-guidance}{https://www.gov.uk/guidance/equality-act-2010-guidance}} these are identified as {\em protected characteristics}, and UK law states that individuals cannot be discriminated against on the basis of these. These protected characteristics typically define subgroups of the population who have historically been disadvantaged in particular situations, such as age discrimination in the workplace, unequal access to healthcare or barriers in education for people with disabilities and gender disparities in political representation, among others.
Driven by the bias that often exists in the training data as a result of these systemic inequalities, 
machine learning approaches often produce biased results (e.g., discrimination in credit market~\cite{Fuster2022} or justice~\cite{Green2018, Johndrow2017} algorithms); there is a growing body of work (often referred to as {\em algorithmic fairness}) that aims to 
identify and mitigate such bias by applying a range of fairness metrics that compare the outcomes achieved by what is identified as advantaged and disadvantaged subgroups of the population (see, e.g., \cite{Hutchinson2019, Mitchell2021} for a review).  

Taking inspiration from the UK Equality Act 2010, we define the concept of \emph{protected attributes} within a multi-agent system, which are any attributes that have been deemed should not disadvantage an agent in terms of its performance within that system. For example, consider a multi-agent setting that includes both artificial agents in the form of autonomous vehicles and human agents who drive their own cars; we may want to ensure that the human agents are not disadvantaged in such a setting. We adapt the following fairness metrics from the algorithmic fairness literature to our multi-agent setting. 
\begin{itemize}
    \item {\em Demographic parity} --  Agents with and without protected attributes should obtain the same expected rewards.
    \item {\em Counterfactual fairness} --  In both a factual and a counterfactual scenario, where the only difference is whether the protected attributes hold for an agent, agents should obtain the same expected rewards.
    \item {\em Conditional statistical parity} --  Within a group of agents characterised by a legitimate factor influencing rewards, agents with and without protected attributes should obtain the same expected rewards.
\end{itemize}

We are able to evaluate different MAS according to these metrics, with the ultimate aim of designing fairer MAS (for example, by configuring the environment in which agents operate to optimise for fairness). Such an approach is inspired by other works outside MAS, such as designing accessible buildings~\cite{Zallio2021} or safe urban environments~\cite{Thompson2020}. Further studies explore environment configurations to optimise rescue operations and autonomous vehicle planning~\cite{Kozůbek19, Karma2015}. However, none of them deal with fairness. Hence, we hope this research can offer valuable insights into domains beyond MAS.

To summarise, the contributions of this paper are as follows.
 We introduce \emph{protected attributes} to MAS -- characteristics that should not impact an agent's expected rewards, all other things being equal. 
We adapt the concepts of \emph{demographic parity}, \emph{counterfactual fairness} and \emph{conditional statistical parity} from the algorithmic fairness literature to the MAS context. 
The future aim of this work is to use these metrics to evaluate and optimise MAS for fairness. 
\smallskip

\noindent{\bf Motivating example.} 
In future urban environments, we may see vehicles operated by humans and vehicles operated by AI undertaking journeys within the same road network. These human and AI agents navigate city streets to reach their destinations, with the rewards they receive dependent on things like time taken and cost of journey. AI-driven vehicles excel by analysing traffic data in real-time, optimising routes, and communicating with other AI vehicles, providing them with an advantage over the human agents in the system, who are generally less efficient at route optimisation and less well-equipped to coordinate with other road users. To mitigate this advantage of AI agents, we might consider altering the road infrastructure, for example, by providing dedicated lanes for human-controlled vehicles.

\section{Related work}
Fairness has attracted the attention of Game Theory and MAS researchers for decades alongside psychologists and economists~\cite{Fehr1999, Falk2006, DeJong2008A, Nowak2000}. Factors such as the rules that govern the system can influence fairness in MAS. For instance, this can be seen in the Ultimatum Game, where fairness is influenced by the dynamics between proposers and responders~\cite{Rand2013, Cimpeanu2021, Kim2023}. In goods allocation or cake-cutting games, the rules depend on the type of good being allocated, for example, whether they are divisible or indivisible, goods or chores~\cite{Aziz2019, Hosseini2023}, and fairness depends on the distribution of goods among the agents~\cite{Amantidis2023, Procaccia2013, Bu2023, Aziz2023, Li2023}.

Agent behaviour can also influence fairness. Fair behaviours often balance the rewards collected by the community and individuals. For example, Zhang and Shah~\cite{Zhang2014} propose a minimum reward for the worst-performing agent while improving the overall rewards of the whole community of agents. However, fairness and reward optimisation can be in tension, and compromises must be made regarding one of the two sides. Jiang and Lu~\cite{Jiang2019} propose a two-step solution consisting of a single policy for each agent based on fair and optimal rewards, with a controller agent who decides which sub-policies to implement to maximise environmental rewards and fairness. Other works~\cite{Hughes2018, Wang2019, Zimmer2021} implement fair optimisation policies within cooperative multi-agent systems, aiming to integrate individualistic and altruistic behaviours. Grupen {\it et al.}~\cite{Grupen2022} introduce a new measure of team fairness, demonstrating how maximising team rewards in cooperative MAS can lead to unfair outcomes for individual agents.  

In contrast to these works, which do not distinguish agents that may be particularly disadvantaged within a system, we consider fairness across agents who do or do not possess protected attributes. We adapt demographic parity~\cite{Dwork2012, Kusner2017}, counterfactual fairness~\cite{Kusner2017} and conditional statistical parity~\cite{Corbett-Davies2017} fairness metrics from the algorithmic fairness literature to the MAS setting. %Additionally, we analyse how changes in environmental configurations impact the three metrics above, enabling the identification of configurations that enhance fairness.

\section{Preliminaries}
A multi-agent system consists of multiple decision-making agents who act and interact in an environment to achieve their goals. 
A {\it multi-agent system} $\Sys = (E, e_o, Ac, P, At, At^{pr}, \tau)$ is characterised by: 
the set of possible {\it environment states} $E$; 
the starting state $e_0$; 
the set of available {\it actions} that may be performed by an agent in the environment $Ac$ (including a null action); 
a {\it population} $P = \{a_1, \ldots, a_n\}$ of {\it agents}; 
the {\it attributes} $At = \{at_1, \ldots, at_m\}$ available to the agents in $P$;
the {\it protected attributes} $At^{pr} \subset \{at_1, \ldots, at_m\}$;
and the non-deterministic {\it state transformer function} $\tau : E \times Ac_1 \times \ldots \times Ac_n \rightarrow E \times [0,1]$ that specifies the probability distribution over the possible resulting states that can occur when each agent in the population performs an action (where the possible null action reflects that an agent chooses not to act).

An {\em agent} $a_x$ within a multi-agent system $(E, e_o, Ac, P, At, At^{pr}, \tau)$ (where $a_x \in P$) is defined as a tuple $(At_x, Ac_x, \pi_x, \rho_x)$ where:
the {\it attribute evaluation function} $At_x: At \rightarrow \{0,1\}$ specifies which attributes hold true for the agent;
$Ac_x \subseteq Ac$ are the {\it actions available to the agent};
the non-deterministic {\it policy} $\pi_x: E \rightarrow Ac_x \times [0, 1]$ specifies how an agent will act in any given state (represented as a probability distribution over the possible actions);
and the {\it reward function} $\rho_x: E \times E \rightarrow \mathbb{R}$ specifies the reward the agent receives for moving between two states. 

A {\it possible run} within a multi-agent system $\Sys = (E, e_o, Ac, P, At, At^{pr}, \tau)$ (where $P$ consists of $n$ agents) is denoted 
$r = (e_0, (ac_1^1, \ldots, ac_1^n), e_1, \ldots, (ac_j^1, \ldots, ac_j^n), e_j)$
where:
for each $a_x \in P$ and for each $i$ such that $0 < i \leq j$,
$(ac_i^x, p) \in \pi_x(e_{i-1})$ and $p>0$;
and for each $i$ such that $0 \leq i < j$, $(e_{i+1}, p) \in \tau(e_i, (ac_i^1, \ldots, ac_i^n))$ and $p>0$.
The set of {\it all possible runs} within a multi-agent system $\Sys$ is denoted $\mathcal{R}^{\Sys}$.

Let $r = (e_0, (ac_1^1, \ldots, ac_1^n), e_1, \ldots, (ac_j^1, \ldots, ac_j^n), e_j) \in \mathcal{R}^{\Sys}$ %be a possible run of 
where $\Sys = (E, e_o, Ac, P, At, At^{pr}, \tau)$. We can determine the probability $r$ will occur, denoted $p(r \mid \Sys)$,  as follows.
\begin{align*}
p(r \mid \Sys) = \\
& \hspace{-0.5cm} \Biggl( \prod_{i=0}^{j-1} \Biggl( \prod_{x=1}^{n} p_x \,\text{where } (ac_{i+1}^x, p_x) \in \pi_x(e_i) \Biggr) \Biggr) \cdot \Biggl( \prod_{i=0}^{j-1} p_i \,\text{where } (e_{i+1}, p_i) \in \tau(e_i, (ac_{i+1}^1, \ldots, ac_{i+1}^n)) \Biggr)
\end{align*}
For a run $r = (e_0, (ac_1^1, \ldots, ac_1^n), e_1, \ldots, (ac_j^1, \ldots, ac_j^n), e_j)$, the reward achieved by an agent $a_x$ is $Rew(a_x, r) = \sum_{i=1}^{j} \rho_x(e_{i-1}, e_i)$.

\smallskip 

\noindent The {\em expected reward} of an agent $a_x$ within a system $\Sys$, denoted $ExpRew(a_x, \Sys)$, is thus
$ExpRew(a_x, \Sys) = {\displaystyle \sum_{r \in \mathcal{R}^\Sys} Rew(a_x, r) . p(r \mid \Sys)}$.

\medskip

\noindent{\bf Motivating example, continued.}
The city traffic consists of a population of cars, each capable of steering, accelerating or braking. Cars also possess attributes like speed or safety features. Cars are either driven by AI or humans, and we consider being driven by humans to be a protected attribute of cars. AI-driven cars can find optimal paths to reach their destination more efficiently than human-driven ones. If we consider agents reaching a hospital, we can foresee fairness problems as AI-driven cars would be advantaged. When the cars act with a specific probability, the environment changes state. Also, each car obtains a reward when reaching its destination. A car's policy is a decision rule based on the state of the crossroads.
 
\section{Fairness in MAS} 
We define fairness by comparing, in different ways, the rewards gathered by individuals or groups of agents possessing and not possessing protected attributes. We adapt \emph{demographic parity}~\cite{Dwork2012, Kusner2017}, \emph{counterfactual fairness}~\cite{Kusner2017} and \emph{conditional statistical parity}~\cite{Corbett-Davies2017} to MAS.

Demographic parity in MAS is achieved when the expected rewards of agents are not influenced by whether or not they possess protected attributes, all else being equal.

\begin{definition}[\textbf{Demographic Parity}]  \label{Demographic Parity}
Let $\Sys = (E, e_o, Ac, P, At, At^{pr}, \tau)$ be a system and let $at^{pr} \in At^{pr}$ be the protected attribute under consideration. 
Demographic parity is satisfied for $at^{pr}$ in $\Sys$ if and only if:
%for all $at^{pr} \in At^{pr}$,
for all $a_x, a_y \in P$,
if $At_x(at^{pr}) = 1$, $At_y(at^{pr}) = 0$, and for all $at^\prime \in At\setminus \{at^{pr}\}$, $At_x(at^\prime) = At_y(at^\prime)$, then
$ExpRew(a_x, \Sys) = ExpRew(a_y, \Sys)$.

Where demographic parity is not satisfied for a particular protected attribute, we can measure the extent to which this is the case, denoted $DemPar(at^{pr}, \Sys)$, as follows.
\begin{equation}
    DemPar(at^{pr}, \Sys) = \\
    {\displaystyle \sum_{\substack{a_x, a_y \in P \mbox{ such that } At_x(at^{pr}) = 1, \mbox{ }At_y(at^{pr}) = 0, 
    \\ \mbox{ and for all }at^\prime \in At\setminus \{at^{pr}\}, At_x(at^\prime) = At_y(at^\prime)
    }}ExpRew(a_x, \Sys) - ExpRew(a_y, \Sys)}  
\end{equation}
\noindent Note that if demographic parity holds for $at^{pr}$ in $\Sys$ then $DemPar(at^{pr}, \Sys) = 0$.
\end{definition}

Counterfactual fairness in MAS is achieved when the expected rewards of agents remain the same in both a factual and a counterfactual world, where in the latter, we change the protected attribute of the agents while keeping all other elements the same.

\begin{definition} [\textbf{Counterfactual Fairness}]  \label{Counterfactual Fairness}
    Let $\Sys = (E, e_o, Ac, P, At, At^{pr}, \tau)$ be a system 
    where $P = \{(At_1, Ac_1, \pi_1, \rho_1), \ldots, (At_n, Ac_n, \pi_n, \rho_n)\}$,
    and let $at^{pr} \in At^{pr}$ be the protected attribute under consideration.  
    Let $\Sys^\prime =  (E, e_o, Ac, P^\prime, At, At^{pr}, \tau)$ be the counterfactual of $\Sys$ such that 
    $P^\prime = \{(At_1^\prime, Ac_1, \pi_1, \rho_1), \ldots, (At_n^\prime, Ac_n, \pi_n, \rho_n)\}$
    where for all $i$ such that $1 \leq i \leq n$: 
    if $At_i(at^{pr}) = 0$, then $At_i^\prime(at^{pr}) = 1$;
    if $At_i(at^{pr}) = 1$, then $At_i^\prime(at^{pr}) = 0$;
    and for all $at \in At\setminus \{at^{pr}\}$, $At_i(at) = At_i^aprime(at)$.
 Counterfactual fairness is satisfied for $at^{pr}$ in $\Sys$ if and only if:
for all $a_x = (At_x, Ac_x, \pi_x, \rho_x) \in P$,
for all $a_x^\prime = (At_x^\prime, Ac_x, \pi_x, \rho_x) \in P^\prime$,
$ExpRew(a_x, \Sys) = ExpRew(a_x^\prime, \Sys^{\prime})$.
Where counterfactual fairness is not satisfied, we can measure the extent to which this is the case, denoted $CountFair(at^{pr}, \Sys)$, as follows.
\begin{equation}
    CountFair(at^{pr}, \Sys) =
    {\displaystyle \sum_{\substack{a_x \in P \mbox{ such that }At_x(at^{pr}) = 1} }ExpRew(a_x, \Sys) - ExpRew(a_x^{\prime}, \Sys^{\prime})} 
\end{equation}
 
\noindent Note that if counterfactual fairness holds for $at^{pr}$ in $\Sys$ then $CountFair(at^{pr}, \Sys) = 0$.
\end{definition}

Conditional statistical parity in MAS is achieved when the expected rewards of agents are not influenced by whether or not they possess protected attributes when conditioning on a legitimate factor, assuming all other elements are the same. A legitimate factor is an attribute that has been identified as one that may legitimately affect an agent's reward.

\begin{definition}[\textbf{Conditional Statistical Parity}]  \label{Conditional Statistical Parity}
Let $\Sys = (E, e_o, Ac, P, At, At^{pr}, \tau)$ be a system,
let $LF \subseteq (At \setminus At^{pr})$ be the set of legitimate factors, and let $at^{pr} \in At^{pr}$ be the protected attribute under consideration.
Conditional statistical parity is satisfied for $at^{pr}$ with $LF$ in $\Sys$ if and only if:
%for all $lf \in At$
for all $a_x, a_y \in P$,
if $At_x(at^{pr}) = 1$, 
$At_y(at^{pr}) = 0$, 
$At_x(at^{lf}) = At_y(at^{lf}) = 1$ for all $at^{lf} \in LF$,
 and for all $at^\prime \in At\setminus \{at^{pr}\}$, $At_x(at^\prime) = At_y(at^\prime)$, then
$ExpRew(a_x, \Sys) = ExpRew(a_y, \Sys)$.

Where conditional statistical parity is not satisfied, we can measure the extent to which this is the case, denoted $CondSP(at^{pr}, LF, \Sys)$, as follows.
\begin{equation}
    CondSP(at^{pr}, LF, \Sys) = \\
    {\displaystyle \sum_{\substack{a_x, a_y \in P \mbox{ such that } At_x(at^{pr}) = 1 \mbox{, } At_y(at^{pr}) = 0, \\
    At_x(at^{lf}) = At_y(at^{lf}) = 1 \mbox{ for all }at^{lf} \in LF,\\
    \mbox{and for all } at^\prime \in At\setminus \{at^{pr}\}, \mbox{ }At_x(at^\prime) = At_y(at^\prime)}}
    ExpRew(a_x, \Sys) - ExpRew(a_y, \Sys)}  
\end{equation}
\noindent Note that if conditional statistical parity holds for $at^{pr}$ with $LF$ in $\Sys$ then $CondSP(at^{pr}, LF, \Sys) = 0$.
\end{definition}

Conditional statistical parity is demographic parity within subsets of the population characterised by legitimate factors. For example, in algorithmic fairness, such a metric is used to verify whether the probability of predicting re-offence for male and female prisoners is the same for similar age groups, which is the legitimate factor~\cite{Berk2021}.

\smallskip

\noindent{\bf Motivating example, continued.}
In the city traffic example, demographic parity would be achieved if the sum of the expected rewards obtained by AI-driven cars and human-driven cars were equal, all other things being equal. In other words, the protected attribute should not affect the expected rewards gathered by the human-driven cars compared to the AI-driven ones.
Counterfactual fairness is achieved if the sum of the expected rewards of the cars remains the same in both a factual and a counterfactual world, where in the latter, agents possess the protected attribute (i.e., cars are driven by humans) while keeping all other factors constant.
Conditional statistical parity is achieved if the sum of the cars' expected rewards is not influenced by whether or not they possess protected attributes when conditioned on a legitimate factor, e.g., a certain range of speed capacity of the cars, assuming all other elements are the same.

\smallskip 

We can use the metrics above to measure fairness of different systems. Our ultimate goal is to optimise systems for these different fairness measures, for example by adjusting the starting state of the environment, or the way the environment responds to the agents' actions. 

\section{Conclusion and future work}
This paper is a first step towards ensuring that certain sub-groups of agents are not disadvantaged in multi-agent systems. 
We identify {\it protected attributes}, which are characteristics that should not disadvantage an agent in terms of its expected rewards. Inspired by algorithmic fairness, we adapt \emph{demographic parity}, \emph{counterfactual fairness} and \emph{conditional statistical parity} to analyse fairness in MAS.
%Second, we analyse how environmental configurations influence the metrics above, allowing us to identify environmental configurations that enhance fairness.
Our metrics assess fairness from various perspectives in any multi-agent system where expected rewards are applicable. Additional metrics from the algorithmic fairness literature, such as equal opportunity, equalised odds~\cite{Hardt2016}, disparate impact~\cite{Feldman2015}, or other metrics based on causal reasoning~\cite{Kilbertus2017, Nabi2018} could be adapted to this setting to capture other aspects of fairness. Our methodology applies to MAS, involving both human and AI agents, as motivated by our example. It could also be used to improve the fairness of human societies by modelling these as multi-agent systems and seeing how changes to the system affect the various fairness metrics defined here. 

In future work, we plan to analyse these fairness metrics experimentally in different settings, both competitive and cooperative, to find system configurations that enhance fairness. We will use techniques such as Bayesian optimisation~\cite{Snoek2012}, evolutionary algorithms~\cite{Vikhar2016} and sparse sampling techniques~\cite{Kearns1999} to try to identify system configurations that optimise for the different fairness metrics.

\section*{Acknowledgments}
This work was supported by UK Research and Innovation [grant number EP/S023356/1], in the UKRI Centre for Doctoral Training in Safe and Trusted Artificial Intelligence (\url{www.safeandtrustedai.org}).

\newpage
\bibliography{bibliography}
\end{document}